\documentstyle[11pt,aaspp4]{article}
\input psfig

\newcommand{\lsim}{\mathrel{\hbox{\rlap{\lower.55ex \hbox {$\sim$}}
                   \kern-.3em \raise.4ex \hbox{$<$}}}}

\begin{document}

\title{Direct measurement of the jet geometry in Seyfert galaxies}

\author
  {J. E. Pringle,\altaffilmark{1,2}, R.R.J. Antonucci\altaffilmark{3},
C. J. Clarke\altaffilmark{1}, A. L. Kinney\altaffilmark{1,2},
H. R. Schmitt\altaffilmark{2} and   J. S. Ulvestad\altaffilmark{4}}
\altaffiltext{1}{Institute of Astronomy, University of Cambridge, Madingley
Road, Cambridge CB3 0HA}
\altaffiltext{2}{Space Telescope Science Institute, 3700 San Martin Drive, 
Baltimore, MD 21218, USA}
\altaffiltext{3}{University of California, Santa Barbara, Physics
Department, Santa Barbara, CA 93106}
\altaffiltext{4}{National Radio Astronomy Observatory, P.O. Box O, 1003 
Lopezville Road, Socorro, NM 87801}

\date{\today}

\begin{abstract}

We demonstrate that, by combining optical, radio and X--ray
observations of a Seyfert, it is possible to provide a direct
measurement of the angle $\beta$ between the direction of the radio
jet and the normal to the plane of the spiral host galaxy. To do so,
we make the assumptions that the inner radio jet is perpendicular to
the X--ray observed inner accretion disk, and that the observed jet
(or the stronger component, if the jet is two-sided) is physically
closer to Earth than the plane of the galaxy. We draw attention to the
possibility of measurement producing a result which is not
self-consistent, in which case for that galaxy, one of the assumptions
must fail.

\end{abstract}


\section {Introduction}

It might be expected, on grounds of symmetry and simplicity, that the
jets emanating from a Seyfert nucleus would emerge at right angles to
the disk of the host spiral galaxy. However, investigations of the
observed {\it distribution} of the angle $\delta$, the difference
between the major axis of the galaxy and the radio jet (seen in
projection on the plane of the sky) indicate that this is not the case
(Ulvestad \& Wilson 1984; Brindle et al. 1990; Baum et al. 1993;
Schmitt et al. 1997). Clarke, Kinney \& Pringle (1998; see also Kinney
et al. 2000) have demonstrated that a reliable estimate of the {\it
distribution} of the angle $\beta$ between the jet axis and the normal
to the galaxy plane can be obtained by considering, for each galaxy in
the sample, the pair of values of $\delta$ and $i$, where $i$ is the
inclination of the galaxy to the line of sight (see Figure 1). They
conclude that the directions of the radio jets are consistent with
being completely uncorrelated with the planes of the host galaxies (see
also Nagar \& Wilson 1999).

The study of this effect, which we suppose can be due to radiation
induced warping of the accretion disk, or interactions of the host
galaxy, is important for the understanding of the inner workings of
Seyfert galaxies. Because we can only observe the galaxy and
jet in projection on the plane of the sky, the measurements of $\delta$
and $i$ merely restrict the jet axis to lie on the plane of a great
circle, centered on the host nucleus, and do not provide a unique
answer for $\beta$ for each galaxy (Section~\ref{geometry}). However,
it has been shown (Nandra et al. 1997; Turner et al. 1998; Weaver \&
Reynolds 1998) that analysis of the shape of the Fe K$\alpha$ line
observed in the X--rays, can yield information on the inclination to
the line of sight, $i_{\rm disk}$, of the inner accretion disk close to
the central black hole, where the line is produced. We note here, that
if one makes the additional assumption that the jet axis is
perpendicular to the inner accretion disk then by measuring the three
angles $\delta$, $i$ and $i_{\rm disk}$ it is possible to obtain a
direct estimate of the angle $\beta$ {\it for individual galaxies}.

\section{Method for Determination of $\beta$}

\subsection{Geometry}
\label{geometry}

For each galaxy we shall assume that we can determine three
observational parameters, $i$, $\delta$ and $\phi = \pm i_{\rm disk}$.
The angle $i$ is the inclination of the plane of the galaxy to the
plane of the sky, or equivalently the angle between the line of sight
and the vector normal to the galaxy plane. The angle $i$ lies in the
range $0^{\circ} < i <90^{\circ}$. We use a Cartesian coordinate system
OXYZ based on the galaxy (see Clarke et al. 1998, and Figure 1).  OX
lies in the disk plane along the apparent major axis, OY lies in the
disk along the apparent minor axis, and thus OZ is the vector normal to
the galaxy plane. In these coordinates the unit vector in the direction
of the line of sight is
\begin{equation}
{\bf k}_s = ( 0, -\sin i, \cos i). 
\end{equation}

The angle $\delta$ corresponds to the difference between the position angle
of the apparent major axis of the galaxy and the direction of the
radio jet projected onto the plane of the sky. By convention
$\delta$ is taken to lie in the range $0^{\circ} < \delta <90^{\circ}$.

For a given pair of values of $\delta$ and $i$, the direction of the
jet, which we denote by a unit vector ${\bf k}_j$, is determined to lie
on a great circle drawn on a unit sphere centered at the origin of our
coordinate system. In the OXYZ coordinates described above the great
circle is the set of points:
\begin{equation}
\label{kjet}
{\bf k}_j     = ( \cos \delta  \sin \phi,
\sin \delta  \cos i \sin \phi - \sin i \cos \phi, 
\sin \delta \sin i \sin \phi + \cos i \cos \phi), 
\end{equation}

Here $\phi$ is the angle between the jet and the line of sight, that
is, it is the angle between the vectors ${\bf k}_s$ and ${\bf k}_j$.
Formally $\phi$ lies in the range $-180^{\circ} < \phi < 180^{\circ}$.
\footnote{We should also note that there is a mirror symmetry to the
problem about the apparent minor axis of the galaxy, the OYZ plane,
which means that reversing the direction of the OX axis leaves the
problem unchanged. Thus, the sign of $k_{jx}$ is not an observationally
meaningful quantity, the jet vector in fact lies on one of two great
circles which are reflections of each other in the OYZ plane, and we
consider only one of them.}

We can make the additional assumption that the observed jet (or the
dominant jet, in the case of a two-sided jet) is the one that lies
above the disk plane of the host galaxy (as seen from Earth), on the
grounds that (apart from relativistic effects) free-free absorption in
the galaxy plane and accretion disk will tend to obscure the
counter-jet (e.g. Pedlar et al., 1998; see also Ulvestad et al., 1999).
\footnote{There is still ambiguity in our knowledge of the relative
position angles; here we assume that the dominant radio jet is between
us and the plane of the galaxy, but we still may not know if the jet is
projected against the half of the galaxy which is nearest to us or if
it is projected against the half of the galaxy which is furthest from
us.}
This can also be determined from observations of HI 21cm absorption by
the galaxy disk. In this case, the jet vector ${\bf k}_j$ is
restricted to lying on one half of the great circle, and the full range
of values of $\phi$ is not permitted. If we define $\beta$ as the angle
the jet vector ${\bf k}_j$ makes with the disk normal OZ (or ${\bf
\hat{z}}$), then we see from Equation~\ref{kjet} that
\begin{equation}
\label{betadet}
  \cos \beta = {\bf k}_j {\bf \cdot \hat{z}} 
= \sin \delta \sin i \sin
\phi + \cos i \cos \phi.
\end{equation}
Then the only permitted values of $\phi$ are those which give $0^{\circ} <
\beta < 90^{\circ}$, or $\cos \beta > 0$. In terms of $\phi$, this means
that $\phi$ lies in the range $\phi_1 < \phi < \phi_1 +180^{\circ}$, where
$\phi_1 \, (<0)$ is the value of
\begin{equation}
\label{phi1}
\phi_1 = \tan^{-1} ( - \cot i/ \sin \delta ),
\end{equation}
which lies in the range $-90^{\circ} < \phi_1 <0^{\circ}$. We note that
physically, if $\phi < 0$, then the jet vector is projected against
the half of the galaxy disk which is nearer to us, whereas $\phi > 0$
corresponds to the jet being projected against the half of the galaxy
disk which is further from us.

\subsection{Measurement of $\beta$}

If, for a given galaxy, we have measurements of $i$ (from the apparent
ellipticity of the galaxy disk on the sky or from the fitting of
kinematical data), of $\delta$ (from radio observations of the
innermost jet structure) and of $i_{\rm disk} = \pm \phi$ (from
modeling of the Fe K$\alpha$ line from X-ray data, or by other means
such as using H$_2$O maser information), then we can determine from
Equation~\ref{betadet} two possible values for $\beta$, one for each
sign of $\phi$. The ambiguity between these values can be resolved by
determining whether the observed jet lies in front of the near side
($\phi < 0$) or the far side ($\phi > 0$) of the galaxy. For most of
the galaxies discussed here, inspection of the optical image shows
unambiguously which side of the minor axis is nearer to us.  The
nearer side has distinct, dark dust lanes while the dust lanes on the
further side cannot be seen because of the intervening bulge (Schmitt
et al. 2000).  When the dust lanes do not leave a distinct signature,
we use the rotation curve of the galaxy together with the assumption
that the spiral arms are trailing.

We give below some examples for which relatively well
determined values of $i, \delta$ and $i_{\rm disk}$ are available.

\subsubsection{Mrk 766 (NGC 4253)}

This is a Seyfert 1 galaxy for which Kinney
et al. (2000) find that $i = 31^{\circ}$ and $\delta =
35^{\circ} \pm 5^{\circ}$. Using ASCA observations of the Fe K$\alpha$
line, Nandra et al (1997) have determined the inner disk inclination,
$\pm \phi$, to be $i_{\rm disk} = 34^{\circ} \pm 3^{\circ}$, assuming
that the disk is around a Schwarzschild black hole, or
$i_{\rm disk} = {36}^{\circ} \: ^{+8}_{-7}$ for a maximal Kerr black hole.
For simplicity we shall just make use of the values for
the Schwarzschild black hole.
Inspection of the HST image of Mrk 766 (Malkan et al. 1998) suggests
that the radio jet (position angle $32^{\circ}$, Nagar et al. 1999) is
seen in projection against the near side of the galaxy. Using the 
values given above together with this knowledge of orientation in
Equation~\ref{betadet}, we conclude that $\beta \simeq 57^{\circ}$.

\subsubsection{NGC 2110}

This is a Seyfert 2 galaxy for which Nagar \&
Wilson (1999) found $i = 43^{\circ}$ and $\delta = 29^{\circ}$. The ASCA
Fe K$\alpha$ data was analyzed initially by Turner et al (1998) who
found that $i_{\rm disk} = 15^{\circ} \: ^{+9}_{-7}$. It was then reanalyzed
by Weaver \& Reynolds (1998) who consider an alternative model for the
line profile and find that (for a Schwarzschild black hole) $i_{\rm
disk}= 50^{\circ} \: ^{+3}_{-4}$. On purely theoretical grounds, the results
derived by Weaver \& Reynolds (1998) are to be preferred, since they
find values of $i_{\rm disk}$ for the four Seyfert 2 galaxies they
analyze in the range $40^{\circ} \lsim i_{\rm disk} \lsim 50^{\circ}$. This is
what one might expect for X--ray visible Seyfert 2 galaxies for which
the line of sight to the nucleus must graze the edge of the
hypothetical molecular torus. Thus we shall adopt $ \phi = \pm 50^{\circ}$.

Using the radio position angle of $10^{\circ}$ (Nagar \& Wilson, 1999)
we find by inspection of the HST image (Malkan et al. 1998) that the
west side of the galaxy is closer to us.  Note that although the radio
source is a triple at the scale of about 100pc, it is evident that the
northern lobe is the stronger on both 100 pc scales and 1 pc scales
(Mulchaey et al. 1994).  We conclude that the jet is seen in projection
against the far side of the galaxy.  Combined with the values given
above, we derive $\beta \simeq 44^{\circ}$.

\subsubsection{NGC3227}
 
This is a Seyfert 1 galaxy for which Kukula et al. (1995) found that
the radio jet is along the position angle $-7^{\circ}$. The parameters
determined by Mundell et al. (1995) using kinematical data are
$i=56^{\circ}$ and $\delta=15^{\circ}$. Nandra et al. (1997) found that the
Fe~K$\alpha$ line profile could be fitted by an inner disk inclination
($\pm\phi$) of $i_{\rm disk}=20^{\circ} \: ^{+10}_{-10}$.
Using the rotation curve of this galaxy and assuming that the spiral
arms are trailing we conclude that the east side of the galaxy is the
nearer side, so the jet is projected against the nearer half of the
galaxy.  Using these values and Equation 3 we find
that $\beta \simeq 63^{\circ}$.  Notice that the accretion
disk inclination found by Nandra et al. (1997) has large uncertainty
so that our results should be taken with caution.

\subsubsection{NGC3516}

This is a Seyfert 1 galaxy for which Nagar et al. (1999) found that
the radio jet is along the position angle 10$^{\circ}$. The parameters
derived by Kinney et al. (2000) are $i=37^{\circ}$ and
$\delta=46^{\circ}$.  Nandra et al. (1999) found that the Fe~K$\alpha$
line could be fitted by an inner disk inclination ($\pm\phi$) of
$i_{\rm disk}= 35^{\circ} \: ^{+1}_{-2}$. The inspection of the HST image
(Malkan et al. 1998) shows that the north side of the galaxy is closer, so
the jet is projected against the nearer side of the galaxy.
Applying Equation 3 to this data we find that $\beta \simeq
66^{\circ}$.

\subsubsection{NGC4051}

This Seyfert 1 galaxy was observed in the radio by Ulvestad \& Wilson
(1984), and has a radio jet along the position angle 81$^{\circ}$.
Lizst \& Dickey (1995) found from kinematical data $i=39^{\circ}$ and
$\delta=51^{\circ}$.  Fitting the Fe K$\alpha$ ASCA spectrum, Nandra et
al. (1997) found $i_{\rm disk}=33^{\circ} \: ^{+5}_{-13}$.  Using the
rotation curve of the galaxy and assuming that the spiral arms are
trailing, we find that the south west side of the galaxy is the nearer
side so that the jet is projected against the side of the galaxy
further from us.  Using these values and Equation 3 we find that $\beta
\simeq 23^{\circ}$.  Note again that the errors in the determination of
$i_{\rm disk}$ are large so that the results should be taken with
caution.

\subsubsection{NGC4258}

This is a Seyfert 2 galaxy for which Miyoshi et al. (1995) measured,
using H$_2$O masers, the rotation curve of a circumnuclear
disk with inner radius of 0.13pc.  According to Miyoshi et al. (1995) , this
disk is at an inclination relative to the line of sight $(\pm\phi)$ of
$i_{\rm disk} = 83^{\circ} \: ^{+4}_{-4}$. The inclination of the host
galaxy is $i=72^{\circ}$, and the major axis position angle is
$148^{\circ}$, determined by van Albada (1980) using HI kinematical
data. Herrnstein et al. (1997) observed that the radio jet lies along
the position angle 0$^{\circ}$, which gives $\delta=32^{\circ}$.  We
find by inspection of the Digitized Sky Survey image that the west
side of the galaxy is the nearer one, so the jet is projected against 
the farther side of the galaxy.  Using Equation 3 and these
values, $\beta \simeq 57^{\circ}$.

Note that we have used the disk inclination derived from water maser
measurements. However, the innermost ring at which the maser
measurements are possible corresponds to several thousand
Schwarzschild radii.  We should bear in mind that the
very inner disk at a few Schwarzschild radii (where the Fe K$\alpha$
is emitted) might not be aligned with the disk at larger radii.

\subsubsection{NGC5506}

This Seyfert 2 galaxy was studied by Kinney et al. (2000), who found
that the radio jet is extended along position angle 70$^{\circ}$,
$i=82^{\circ}$ and $\delta=21^{\circ}$.  According to Wang et
al. (1999) the Fe K$\alpha$ line of this galaxy can be fitted by an
accretion disk with $i_{\rm disk}=40^{\circ} \: ^{+10}_{-10}$. From
inspection of the HST image (Malkan et al. 1998) we find that the
south side is the nearer side, so the jet is projected against the
farther side of the galaxy.  Using Equation 3 we find that
$\beta \simeq 70^{\circ}$.

\section{Discussion}

We have shown that measurement of the three parameters $\delta$, the
projected angle between the galaxy major axis and the radio jet, $i$,
the inclination of the galaxy disk, and $i_{\rm disk}$, the
inclination of the inner accretion disk, together with the assumption
that the radio jet is perpendicular to the inner accretion disk,
yields (Equation~\ref{betadet}) two possible values of the angle
$\beta$, the angle between the radio jet and the plane of the host
galaxy.  In addition, the ambiguity between the two possible values of
$\beta$ can be resolved if one makes the additional assumption that the
observed radio jet (for a one-sided jet, or in the case of a two-sided
jet, for the stronger component) is closer to Earth than the
host galaxy plane, and if one is able to determine whether the jet is
seen in projection against the near side or the far side of the
galaxy.

It is of some interest, however, that the method we have described
above can fail.  If the radio jet is seen to lie in projection against
the nearer side of the galaxy to the Earth, then for a given pair of
values of $i$ and $\delta$, the value of $\phi(<0)$ must lie in the
range $\phi_1 \leq \phi \leq 0$, where $\phi_1$ is given by
Equation~\ref{phi1}. However, if, for such a galaxy, the value of the
inner disk inclination $i_{\rm disk}$, derived from analysis of the
X--ray data were to exceed $| \phi_1 |$, we would have an
inconsistency. This would imply that one (or more) of our assumptions
had failed. Thus an important observational test of the assumptions
underlying this analysis is to look for galaxies which might give rise
to such a contradiction. Conversely, if, after measurements of a large
sample of galaxies, none is found to show this inconsistency, we may
begin to have confidence in the assumptions and the method.
For the 7 galaxies measured for this paper, no inconsistencies have been
found.

For those targets for which optical and radio data (i.e. $\delta$ and
$i$) are already available, the most fruitful ones to pursue with
X--ray observations in a search for self-consistency would be those
galaxies with large inclinations $i$, with large values of $\delta$,
and (of course) whose radio jets lie in projection against the near
side of the galaxy. An example of such a galaxy is the Seyfert 2, NGC
4388, for which $i = 70^{\circ}$, $\delta = 70^{\circ}$ (Kinney et
al. 2000), and for which the jet is seen in projection against the near side of
the galaxy. From Equation~\ref{phi1} we see that $\phi_1 = -
21^{\circ}$.  The galaxy was detected by ASCA (Iwasawa et al. 1997)
but due to the large column density ($N_H = 4 \times 10^{23} cm^{-2}$)
they detected only a narrow, gaussian shaped Fe K$\alpha$ line and
thus were not able to determine the accretion disk orientation (XMM-quality
data may still be able to determine it). If we
find this galaxy, as for the Seyfert 2s analyzed by Weaver \& Reynolds
(1998) has $i_{\rm disk} \simeq 40^{\circ}$ -- $50^{\circ} > |\phi_1|$, we
would have an inconsistency. However, for this galaxy, it might be
that $i_{\rm disk} < 21^{\circ}$, and that it appears to be a Seyfert
2 because of the strong dust absorption seen in this nearly edge on
galaxy, rather than because the inner disk (and molecular torus) is at
a large angle to the line of sight. This is
consistent with the detection of a gaussian Fe K$\alpha$ line.

In a search for objects which might show inconsistency, the only
targets for which X--ray analysis is already available, which are
worth pursuing with radio observations, are those objects for which
the sum of the measured inclinations of the galaxy disk and the inner
accretion disk exceed $90^{\circ}$, that is for which $i + i_{\rm
disk} > 90^{\circ}$. For such galaxies, the inconsistency arises if
the radio jet is observed to lie in projection against the near side
of the galaxy, and if $\delta > \delta_{\rm max}$, where
\begin{equation}
\sin \delta_{\rm max} = \cot i \, \cot i_{\rm disk}
\end{equation}.

We identify here two such galaxies, both Seyfert 2s, whose inner disk
inclinations have been determined by Weaver \& Reynolds (1998). They
are:

$\bullet$ MCG -5-23-16, for which $i_{\rm disk} = 52^{\circ}$, $i =
63^{\circ}$ (de Vaucouleurs et al. 1991), and for which we find
$\delta_{\rm max} = 23^{\circ}$,

and 

$\bullet$ NGC 7314, for which $i_{\rm disk} = 46^{\circ}$, $i = 63^{\circ}$
(de Vaucouleurs et al. 1991), and for which we find 
$\delta_{\rm max} = 29^{\circ}$.

Both galaxies are unresolved by VLA (Ulvestad \& Wilson 1984, Ulvestad,
unpublished); sensitive VLBI observations will probably be necessary
to determine a radio position angle.

\section{Conclusion}

We have described a method for determining the angle $\beta$ the jet
in a Seyfert galaxy makes with the normal to the host galaxy plane,
and have applied the method to seven systems. The results are
summarized in Table 1. \footnote{Given the quality, and the
heterogeneous nature of the current data, we have not attempted a
formal error analysis.} We note that the four Seyfert 1 galaxies have
values for $\beta$ of 23$^{\circ}$, 62$^{\circ}$, 63$^{\circ}$ and
66$^{\circ}$, while the three Seyfert 2 galaxies have values of
44$^{\circ}$, 57$^{\circ}$, and 70$^{\circ}$.  The fact that no values
are near $0\arcdeg$ is consistent with previous conclusions that
radio jets in Seyferts do not tend to align with the galaxy axes.
We have also drawn
attention to the method being able to give rise to an inconsistency,
in which case we would be able to conclude {\em either} that the
observed jet is on the far side of the host galaxy plane (as seen from
Earth), {\em or} that the jet is not emitted perpendicularly to the
accretion disk. 

 The launch of the X--ray observatories XMM and Chandra, coupled with
optical and radio observations, should enable measurements of $\beta$
to be made for a much larger sample of galaxies and with greater
accuracy. This will open up the possibility for a better understanding
of the geometry of the accretion processes around the active nucleus,
and of the processes which give rise to the misalignment between
galaxy disks and radio jets/inner accretion disks.

\acknowledgements JEP is grateful for continued support from the
STScI visitor program.  ALK would like to thank IoA for support under
its visitor program.  This work was supported by NASA grants
NAGW-3757, AR-5810.01-94A, AR-6389.01-94A and the HST Director
Discretionary fund D0001.82223.

\begin{deluxetable}{lcrrlr}
\tablewidth{0pc}
\tablecaption{Measurements and Results}
\tablehead{\colhead{Name}&\colhead{Type}&\colhead{$\delta$}&
\colhead{$i$}&\colhead{$i_{disk}=|\phi|$}&\colhead{$\beta$}}
\startdata
MRK766&  1&35$^{\circ}$&31$^{\circ}$&34$^{\circ}\pm3^{\circ}$&62$^{\circ}$\cr
NGC3227& 1&15$^{\circ}$&56$^{\circ}$&20$^{\circ}\pm10^{\circ}$&63$^{\circ}$\cr
NGC3516& 1&46$^{\circ}$&37$^{\circ}$&35$^{\circ +1}_{~ -2}$&66$^{\circ}$\cr
NGC4051& 1&51$^{\circ}$&39$^{\circ}$&33$^{\circ +5}_{~ -13}$&23$^{\circ}$\cr
NGC2110& 2&29$^{\circ}$&43$^{\circ}$&50$^{\circ +3}_{~ -4}$&44$^{\circ}$\cr
NGC4258& 2&32$^{\circ}$&72$^{\circ}$&83$^{\circ}\pm4^{\circ}$&57$^{\circ}$\cr
NGC5506& 2&21$^{\circ}$&82$^{\circ}$&40$^{\circ}\pm10^{\circ}$&70$^{\circ}$\cr
\tablenotetext{}{}
\enddata
\end{deluxetable}

\begin{figure}
\psfig{figure=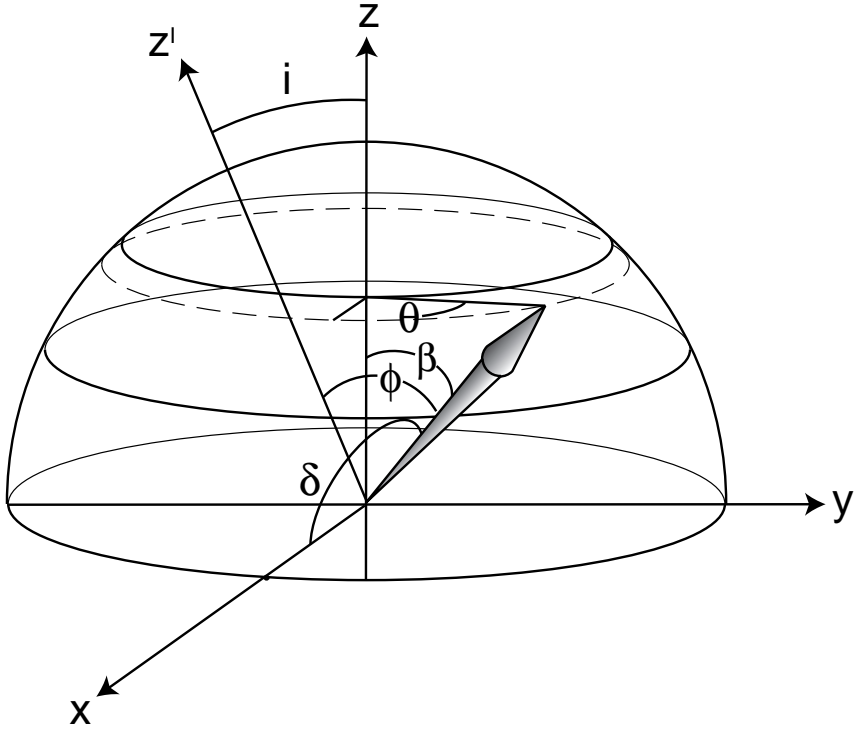,height=10cm}
\caption{
The galaxy lies in the XY plane, with coordinates placed so that the
apparent major axis is the X axis and the galaxy axis is Z.  The line of 
sight Z$^{\prime}$, designed by the vector ${\bf k}_s$, is in the plane of 
the paper. The angle of inclination is $i$.  The position angle between the 
apparent major axis of the galaxy and the radio jet projected onto the sky 
plane is $\delta$.  The radio jet, whose vector is given as 
${\bf k}_j$, is designated by an arrow.  The angle between the radio
jet and the galaxy axis is $\beta$.  The angle between the line of sight
and the radio axis, commonly referred to as the opening angle of the
active galaxy, is $\phi$.  For an accretion disk 
perpendicular to the jet, $\phi = \pm i_{\rm disk}$.
}
\end{figure}
\end{document}